\documentclass[pdflatex,sn-mathphys-num]{sn-jnl}

\usepackage{graphicx}
\usepackage{multirow}
\usepackage{subcaption}
\usepackage{amsmath,amssymb,amsfonts}
\usepackage{amsthm}
\usepackage{mathrsfs}
\usepackage{xcolor}
\usepackage{textcomp}
\usepackage{manyfoot}
\usepackage{booktabs}
\usepackage{algorithm}
\usepackage{algorithmicx}
\usepackage{algpseudocode}
\usepackage{listings}
\usepackage{hyperref}
\usepackage{lineno}
\usepackage{caption}
\usepackage{xspace}
\usepackage{titlesec}
\usepackage{float}
\usepackage{siunitx}
\usepackage{comment}

\newcommand{\antideuteron}{\ensuremath{\overline{\mathrm{d}}}\xspace}
\newcommand{\hethreebar}{\ensuremath{{}^{3}\overline{\mathrm{He}}}\xspace}
\newcommand{\hefourbar}{\ensuremath{{}^{4}\overline{\mathrm{He}}}\xspace}
\newcommand{\neutralino}{\ensuremath{\chi}\xspace}
\newcommand{\neutralinobar}{\ensuremath{\overline{\neutralino}}\xspace}
\newcommand{\chichibar}{\ensuremath{\neutralino\neutralinobar}\xspace}

\newcommand{\Lbbar}{\ensuremath{\overline{\Lambda}_{b}}\xspace}

\newcommand{\antibHadron}{\ensuremath{\overline{\rm H}_b}\xspace}
\newcommand{\Bminus}{\ensuremath{\mathrm{B}^{-}}\xspace}

\newcommand{\gevc}{\ensuremath{{\rm GeV}/c}\xspace}
\newcommand{\pT}{\ensuremath{\rm{p_{\mathrm{T}}}}\xspace}

\newcommand{\reactionLbtod}{\ensuremath{\Lbbar \to \overline{\mathrm{d}} + X}\xspace}
\newcommand{\brLbtod}{\ensuremath{\mathrm{BR}\!\left(\reactionLbtod \right)}\xspace}
\newcommand{\reactionBtod}{\ensuremath{\Bminus \to \overline{\mathrm{d}} + X}\xspace}
\newcommand{\brBtod}{\ensuremath{\mathrm{BR}\!\left(\reactionBtod\right)}\xspace}
\newcommand{\bbbar}{\ensuremath{b\overline{b}}\xspace}
\newcommand{\gevctwo}{\ensuremath{{\rm GeV}/c^2}\xspace}
\newcommand{\bquark}{\ensuremath{b}\xspace}
\newcommand{\bbarquark}{\ensuremath{\overline{\bquark}}\xspace}

\newcommand{\Fig}[1]{\ensuremath{\mathrm{Fig.}~\ref{#1}}\xspace}
\newcommand{\Tab}[1]{\ensuremath{\mathrm{Tab.}~\ref{#1}}\xspace}
\newcommand{\Eq}[1]{\ensuremath{\mathrm{Eq.}~\ref{#1}}\xspace}

\newcommand{\firsttune}{\texttt{M017}\xspace}
\newcommand{\secondtune}{\texttt{M023}\xspace}
\newcommand{\thirdtune}{\texttt{CRmode2}\xspace}
\newcommand{\probQQtoQ}{\ensuremath{\texttt{probQQtoQ}}\xspace}
\newcommand{\PYTHIA}{\texttt{PYTHIA}\xspace}
\newcommand{\momCoal}{\ensuremath{\rm{p_{\rm{coal}}}}\xspace}
\newcommand{\Argonne}{\textit{Argonne}\xspace}
\newcommand{\vEighteen}{\ensuremath{v_{18}}\xspace}
\newcommand{\ArgonneVEighteen}{\Argonne $\vEighteen$\xspace}
\newcommand{\proton}{\ensuremath{\rm{p}}\xspace}
\newcommand{\neutron}{\ensuremath{\rm{n}}\xspace}
\newcommand{\antiproton}{\ensuremath{\overline{\proton}}\xspace}
\newcommand{\antineutron}{\ensuremath{\overline{\neutron}}\xspace}
\begin{document}

\title[]{Antideuteron production from beauty-hadron decays: a first phenomenological study}

\author*[1,2]{\fnm{Marta} \sur{Razza}}\email{marta.razza2@unibo.it}
\author[1,2]{\fnm{Francesca} \sur{Bellini}}\email{f.bellini@unibo.it}
\author[3,4]{\fnm{Nicol\'o} \sur{Jacazio}}\email{nicolo.jacazio@cern.ch}

\affil*[1]{\orgdiv{Department of Physics and Astronomy A. Righi},
  \orgname{University of Bologna},
  \orgaddress{\street{Via Irnerio 46}, \city{Bologna},
    \postcode{40126}, \state{Italy}}}

\affil*[2]{\orgdiv{INFN, Division of Bologna},
  \orgaddress{\street{Via Berti Pichat 6/2}, \city{Bologna},
    \postcode{40127}, \state{Italy}}}

\affil[3]{\orgdiv{Department of Science and Technological Innovation},
  \orgname{University of Piemonte Orientale},
  \orgaddress{\street{Via T. Michel, 11}, \city{Alessandria},
    \postcode{15121}, \state{Italy}}}

\affil[4]{\orgdiv{INFN, Division of Turin},
  \orgaddress{\street{Via Pietro Giuria 1}, \city{Turin},
    \postcode{10125}, \state{Italy}}}
\abstract{\unboldmath
Light antinuclei, such as antideuteron (\antideuteron) and antihelium (\hethreebar, \hefourbar), provide a link between collider physics and indirect Dark Matter searches.
Despite extensive studies of antinucleus production in high-energy collisions, \antideuteron production from beauty-hadron decays remains experimentally unconstrained and has not yet been quantitatively predicted.
In this work, we present the first phenomenological study of \antideuteron production from \Lbbar baryon and \Bminus meson decays, providing the first estimates of the corresponding branching ratios.
Beauty-hadron decays are simulated with PYTHIA using realistic input kinematics and three hadronization scenarios. 
Antideuteron formation is modelled through a quantum-mechanical coalescence approach based on an \antideuteron wave function derived from the \ArgonneVEighteen nucleon--nucleon potential.
Depending on the adopted hadronization model, we estimate inclusive branching ratios to be
$(5.68 \pm 0.02)\times10^{-4}<\brLbtod<(1.408 \pm 0.004)\times10^{-3}$ and $(7.4 \pm 0.3)\times10^{-6}<\brBtod<(4.34 \pm 0.07)\times10^{-5}$.
The predicted rapidity- and transverse-momentum-differential yields populate the kinematic region where \antideuteron can be identified by the ALICE experiment, motivating dedicated searches for these decay channels.
These results provide a quantitative benchmark for \antideuteron production from beauty-hadron decays and establish a phenomenological framework to support future experimental searches, with potential implications beyond collider physics.
}
\keywords{Antinuclei, beauty-hadrons, dark matter}
\maketitle
\titleformat{\section}
{\Large\bfseries}{\thesection}{1em}{}[\titlerule]
\section*{Introduction}
Light antinuclei, such as antideuteron and antihelium, are regarded as exceptional messengers in astrophysical searches for exotic antimatter sources. 
The low expected background flux of antinuclei from conventional cosmic-ray interactions with the interstellar medium (ISM) enhances their appeal as a potential signal channels for exotic processes such as dark matter annihilation or decay~\cite{fornengo2013dark,donato2008antideuteron, carlson2014antihelium, cirelli2014anti, Blum:2017qnn}.

Antinuclei have been successfully produced and observed at accelerator experiments, from the discovery of antideuterons at CERN and at Brookhaven National Laboratory (BNL)~\cite{massam1965experimental, dorfan1965observation}, to the first detection of \hefourbar by the STAR Collaboration at BNL~\cite{star2011observation}.
However, no antinucleus with mass number $A\geq 2$ has yet been observed in cosmic rays.
Dedicated searches are currently underway with the Alpha Magnetic Spectrometer (AMS-02)~\cite{tomassetti2015ams}, mounted on the International Space Station and collecting data since May 2011, and the General AntiParticle Spectrometer (GAPS)~\cite{quinn2018gaps}.
The first Antarctic GAPS flight was completed in January 2026, and further flights are planned.

Among the scenarios postulating a particle nature of dark matter, Weakly Interacting Massive Particles represent one of the most theoretically motivated candidates, with the neutralino (\neutralino) arising from supersymmetric extensions of the Standard Model.
The annihilation or decay of such candidates could result in light antinuclei, such as antideuterons and antihelium nuclei (both \hethreebar and \hefourbar), in the final state~\cite{donato2008antideuteron, carlson2014antihelium, cirelli2014anti}.
These are subsequently expected to propagate and survive through the Galaxy and the heliosphere before being detected in space-borne or balloon-borne experiments near Earth~\cite{alice2023measurement, Serksnyte:2022onw}.
Antinuclei produced by the interaction of primary cosmic rays (CR) with the ISM, primarily via the proton--proton (pp) and proton--nucleus (pA) channels, known as secondary antinuclei, constitute an irreducible background for the signal of primary antinuclei generated by dark matter annihilations.
At kinetic energies below $\approx 1$~GeV/n, the dark matter signal is predicted to exceed the expected flux of secondary antinuclei by 1-2 orders of magnitude~\cite{korsmeier2018prospects, heisig2024darkraynet,Stefanuto:2026wxy}.

Following the 2018 announcement by AMS-02 of the potential detection of few \hethreebar and \hefourbar events~\cite{ting2018ams,choutko2008antideuteron}, a pioneering study~\cite{winkler2021dark} was proposed to investigate the production of antihelium from the decay of the \Lbbar baryon, formed as a consequence of the hadronisation of the \bbarquark produced by dark matter annihilation.
Due to helicity suppression effects, the preferred channel is $\chichibar \rightarrow \bbbar$, which subsequently hadronizes into particles containing \bquark or \bbarquark quarks.
Among beauty-baryons, the \Lbbar represents an ideal candidate for the production of nucleons and antinucleons due to its large rest mass, around 5.6 \gevctwo, above the 4.7 \gevctwo rest mass of the three antinucleons involved in the formation of \hethreebar plus the two additional nucleons required for baryon number conservation.
Within the Standard Model of particle physics, a detectable antihelium flux would then be given by the process $\Lbbar \longrightarrow \hethreebar +\rm{p} +\rm{p}$.
In~\cite{winkler2021dark}, assuming a branching ratio $\mathcal{B}(\Lbbar \to \hethreebar)\approx 10^{-6}$, the resulting \hethreebar flux was found to be of the order of $O(10^{-7})$, within the 
sensitivity reach of the AMS-02 experiment after 10 years of exposure.
Meanwhile, recent preliminary results by the LHCb collaboration constrained this scenario by setting an upper limit of $\mathcal{B}(\Lbbar \to \hethreebar + \rm{p+p}) < 1.9 \times 10^{-9}$ at 90\% CL \cite{LHCb:2024antihelium}, well below the predicted one.
However, beauty-hadron decays are interesting not only in the context of indirect dark matter searches. They also provide a unique laboratory to investigate light antinucleus formation in heavy-flavour decays and may become experimentally accessible at collider experiments.

In this work, we turn our attention on the \antideuteron production channel. In general, antideuteron production is expected to be larger than antihelium production, as it involves only two antinucleons. At the LHC, for instance, the production of \antideuteron in pp collisions is reported to be $10^3$ times more abundant than that of \hethreebar~\cite{ALICE:2019bnp}.
It is then natural to ask whether \antideuteron production receives any significant contribution from beauty-hadron decays. In this case, both \Lbbar and \Bminus states should be considered.
Unlike \hethreebar production, which is kinematically forbidden in \Bminus decays once baryon-number conservation is imposed, \antideuteron production is allowed.
To our knowledge, no dedicated experimental search for \antideuteron from \Lbbar or \Bminus decays has been reported so far, and no dedicated phenomenological study of these channels appears to be available in the literature.

Therefore, we present a study of the decay of beauty hadrons into antideuterons to obtain a first estimate of the $\Lbbar\rightarrow \antideuteron + \rm{X}$ and $\Bminus\rightarrow \antideuteron + \rm{X}$ branching ratios.
This analysis combines simulations based on the \PYTHIA general-purpose QCD-inspired event generator with a state-of-the-art coalescence model for antinuclei formation.
The paper is organised as follows.
Section~\ref{sec:pythia} describes the \PYTHIA simulation framework, including the simulation setup and the specific generator configurations adopted to model beauty-hadron decays.
Section~\ref{sec:coalescence} presents the model used to describe the formation of the \antideuteron.
Section~\ref{sec:Results} presents the results in terms of the estimated branching ratios for each configuration under study, together with a systematic analysis of the relevant decay channels.
We conclude by discussing how these results may steer future searches for the $\Lbbar\rightarrow \antideuteron + \rm{X}$ and $\Bminus\rightarrow \antideuteron + \rm{X}$ decay channels with accelerator-based experiments.
Our study contributes, on the one hand, to the understanding of antinuclei production mechanisms through the exploration of previously uninvestigated channels.
On the other hand, it may provide valuable input for applications in the indirect dark matter search context.

\section{Simulation with PYTHIA 8.3}
\label{sec:pythia}
To simulate a realistic \pT~distribution of the beauty hadrons of interest, \Lbbar and \Bminus, we employed perturbative QCD calculations at first-order next-to-leading logarithmic accuracy (FONLL)~\cite{cacciari1998pt} for pp collisions at $\sqrt{s} = 13.6$ TeV as input for the \PYTHIA 8.3 event generator~\cite{bierlich2022comprehensive}. FONLL is found to reproduce the centre-of-mass-energy dependence of the total beauty production cross section ($\sigma(\rm{pp\rightarrow b\bar{b}})$) from $\sqrt{s} = 200$ GeV to 13 TeV, the momentum dependence of non-prompt D meson (D$^{0}$, D$^{+}$, and D$_{s}^{+}$ from beauty-hadron decays) production~\cite{ALICE:2024xln}, as well as the beauty-hadron production cross section $\sigma(\rm{pp\rightarrow H_{b}+X})$ and its excitation function~\cite{LHCb:2016qpe, ALICE:2026xuh} in pp collisions at $\sqrt{s} = 13$ TeV.
For this study, the cross-section of beauty hadrons is evaluated in the pseudorapidity range $|\eta| < 1$.

The transition of a beauty quark into a given beauty-hadron species is described by the fragmentation function, which encodes the non-perturbative hadronisation process.
In phenomenological applications, the relative production rates of the different beauty-hadron species are commonly expressed in terms of fragmentation fractions.
These quantities are typically constrained using measurements in $\mathrm{e^+e^-}$ or ep collisions~\cite{Skands:2014pea}, under the assumption that the hadronisation process is universal, namely independent of the collision system.
Measured meson-to-meson production ratios for charm and beauty hadrons in pp and pA collisions at the LHC are consistent with earlier results from $\mathrm{e^+e^-}$ and ep collisions~\cite{Altmann:2024kwx}, supporting the use of fragmentation fractions measured in $\mathrm{e^+e^-}$ collisions for normalising beauty-meson production~\cite{ALICE:2026xuh}.
Accordingly, the \Bminus \pT-distribution (or \textit{spectrum}) is obtained directly from the FONLL output by applying the fragmentation fraction $\rm{f(b\rightarrow B^{-})}$, as measured in $\mathrm{e^+e^-}$ collisions~\cite{navas2024review}.

In contrast, heavy-flavour baryon production in hadronic collisions at LHC energies shows a significant enhancement relative to meson production when compared to the $\mathrm{e^+e^-}$ \pT-independent baseline~\cite{Bala:2025usg, LHCb:2023wbo,acharya2023study}.
The $\rm{f(b\rightarrow\Lambda_b)/f(b\rightarrow B)}$ ratio measured in pp collisions exhibits a \pT-dependence~\cite{LHCb:2011leg,aaij2019measurement}. 
Therefore, following \cite{acharya2023study}, the \Lbbar distribution is derived from the previously obtained FONLL \Bminus spectrum by multiplying for the most recent \pT-dependent measurement of the $f_{\Lambda_b}/(f_u + f_d)$ ratio by LHCb at $\sqrt{s} = 13\;\mathrm{TeV}$~\cite{aaij2019measurement},
\begin{equation}
  \frac{f_{\Lambda_b}}{f_u + f_d}(\pT)
  \;=\; A\bigl[p_1 + \exp\;\bigl(p_2 + p_3\,\pT\bigr)\bigr],
  \label{transitionrateLHCb}
\end{equation}
where the beauty-hadron transverse momentum is $\pT = \pT(\rm{H_b})$, and the parameters are $\rm{A}=1\pm 0.061$, $\rm{p_1} = (7.93 \pm 1.41) \times 10^{-2}$, $\rm{p_2} = -1.022 \pm 0.047$, and \mbox{$\rm{p_3} = -(0.107 \pm 0.002)\;\mathrm{GeV}^{-1}$}.

The subsequent decays of the \Lbbar and \Bminus hadrons are handled entirely by \PYTHIA, which simulates the full decay chain according to its decay tables and tunes.
The $\PYTHIA$ tunes considered in this work correspond to three different configurations, selected to span different hadronization scenarios, ranging from a tune optimized for $\mathrm{e}^{+}\mathrm{e}^{-}$ data (\firsttune), to one tuned to pp beauty-baryon production (\secondtune), and a color-reconnection model (\thirdtune).

The choice labelled \firsttune follows the study presented in~\cite{DiMauro:2025vxp}, which aims at reproducing the most recent measurements of the transition rates $f(\mathrm{b}\!\to\! \Lbbar,\,\mathrm{B^+})$ in $\mathrm{e^+e^-}$ collisions as reported by the Heavy Flavor Averaging Group (HFLAV)~\cite{amhis2021averages}. 
This is based on the Monash2013 tune~\cite{Skands:2014pea} with a modified value of \probQQtoQ.
The \probQQtoQ parameter modifies the flavour dynamics in the Lund string fragmentation model.
Specifically, it adjusts the probability for the production of a diquark-antidiquark pair from the string, thereby influencing the relative rates of baryon and meson production during hadronisation~\cite{10.21468/SciPostPhysCodeb.8} thus affecting the number of nucleons available in the final state.

The second chosen tune, \secondtune, starts from the same idea as the first but was developed specifically for this work and is based on variations in the \probQQtoQ parameter.
The tune attempts to reproduce the ratio in \Eq{transitionrateLHCb} as a function of \pT in pp collisions at $\sqrt{s}=13.6\;\rm{TeV}$ by performing a scan of the \probQQtoQ parameter in the range $[0.17,\,0.25]$.
The configuration with the lowest $\chi^2$ corresponds to $ \probQQtoQ = 0.23$.
We note that even for this value, the simulation still exhibits a tension with the \pT-dependent trend described by \Eq{transitionrateLHCb} in the low momentum region.

\begin{figure}[H]
  \centering
  \includegraphics[width=0.8\linewidth]{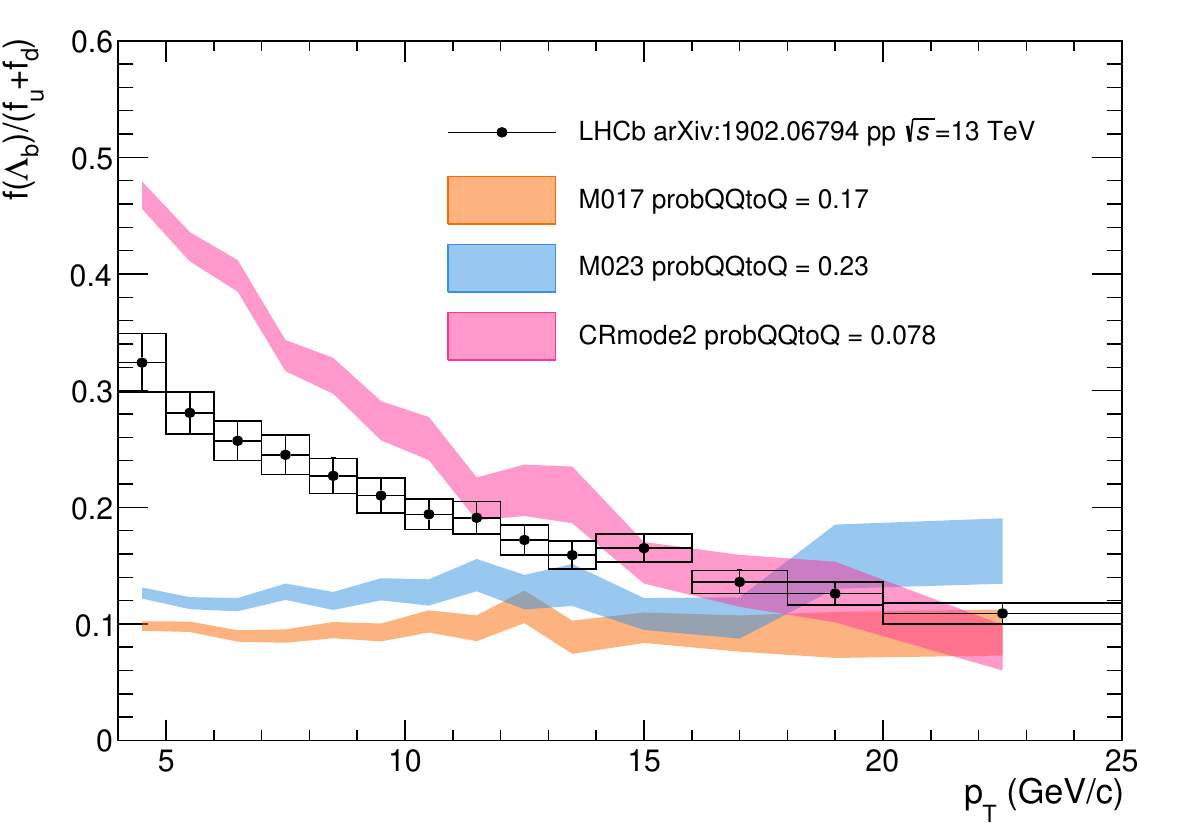}
  \caption{Comparison between the three different configurations used in \PYTHIA and the LHCb data for the $\rm{H_b}$ fractions~\cite{aaij2019measurement}.
    The orange curve corresponds to \firsttune, the Monash2013 tune with $\probQQtoQ = 0.17$ to best match
    the $e^+e^-$ measurements.
    The light-blue one, corresponding to \secondtune, uses the Monash2013 tune with $\probQQtoQ = 0.23$ to reduce the discrepancy with the data in pp collisions.
    The pink one is extracted using the \thirdtune tune and is characterised by $\probQQtoQ = 0.078$.}
  \label{fig:scanTune}
\end{figure}

The \thirdtune configuration is the QCD-based Colour Reconnection mode 2, with the parameters reported in~\cite{christiansen2015string}.
This model allows colour rearrangements beyond the leading-colour approximation, favouring lower string-length configurations and permitting junction-like topologies during hadronisation.
This tune was selected because such effects have been shown to be important to model baryon production in pp collisions at the LHC and to improve the description of baryon-to-meson ratios~\cite{acharya2023study}, though the level of agreement remains significantly flavour- and \pT-dependent.
Figure~\ref{fig:scanTune} shows the comparison between the $f_{\Lambda_b}/(f_u + f_d)$ predicted under the different tunes and the LHCb data.
The results show that varying the \probQQtoQ parameter while keeping the baseline Monash2013 tune is not sufficient to reproduce the \pT-dependent ratio observed in data.
The current parameter choices provide a description which is quantitatively in agreement with the high-\pT region rather than the low-\pT one.
By contrast, \thirdtune reproduces the \pT-dependence of the $\rm{H_b}$ distributions qualitatively.

In addition to the tune, for the purpose of this study, particular attention was given to the modelling of decays in \PYTHIA.
Decays can be classified as either exclusive or inclusive.
In exclusive decays, the final-state particles are fully specified.
\PYTHIA randomly selects one of the available exclusive channels according to its branching ratio and distributes the final-state particles within the allowed phase space.
Instead, in inclusive decays, the final-state structure is not known a priori and only the statistical features of the decay are constrained.
In such cases, \PYTHIA begins with partonic decays (e.g., $\rm{b} \rightarrow \rm{q}$ or $\rm{b} \rightarrow$ diquark), followed by a parton shower and the hadronisation to produce the final hadronic states.
As stated in the \PYTHIA manual~\cite{10.21468/SciPostPhysCodeb.8}, the \probQQtoQ parameter directly impacts on the decay chains, as it modifies the probability of forming diquarks during hadronisation, therefore affecting the branching ratios of the \Lbbar baryon and the \Bminus meson.
Consequently, both parameter variations and the choice of tune can affect the final-state composition and structure of the decay chain.
A comparison between the three different configurations in terms of decays in which \antibHadron are producing antinucleons is reported in Appendix~\ref{sec:Appendix}.
After the simulation of the particles and their decays, the final products are fed to an afterburner that implements the production of \antideuteron through a coalescence approach.

\section{Antinuclei formation via coalescence}
\label{sec:coalescence}

Early coalescence models~\cite{schwarzschild1963production, butler1963deuterons, kapusta1980mechanisms} relied on a deterministic momentum-proximity criterion, according to which a proton--neutron pair forms a deuteron if its relative momentum in the pair rest frame satisfies the condition $|\vec{p}_{\proton} - \vec{p}_{\neutron}| < \momCoal$, where \momCoal is a fixed threshold called coalescence momentum. 
Subsequent developments introduced a phase-space proximity condition: $\Delta p < \momCoal$ and $\Delta r < r_{\rm{d}}$~\cite{sombun2019deuteron}, therefore adding a constraint also on the spatial separation between two nucleons, with $r_{\rm{d}}$ being a characteristic spatial scale as for instance the deuteron radius.
These approaches, however, neglect some aspects of nucleus production that are inferred from experimental observations: the coalescence probability depends on the \pT per nucleon (\pT/A) of the produced nuclei~\cite{acharya2018production} as well as on the final state charged-particle multiplicity~\cite{ALICE:2019dgz,ALICE:2022veq}.
The multiplicity, in turn, is commonly considered as a proxy for the effective volume of the emission source, which scales approximately with the cube of the source radius measured via momentum correlation (femtoscopy) techniques~\cite{ALICE:2011kmy, ALICE:2015hvw, ALICE:2015tra, ALICE:2019dgz, ALICE:2025wuy}.
As the source volume increases, the typical spatial separation between produced particles increases, reducing the coalescence probability.
Therefore, a realistic modelling of \antideuteron formation should simultaneously account for three key ingredients: the size of the emission source, the spatial extent of the cluster to be formed (i.e. the size of the \antideuteron), and the transverse momentum of the participating nucleons.
A Wigner-function formalism naturally incorporates all of these aspects into the coalescence calculation~\cite{Scheibl:1998tk, Blum:2019suo, Bellini:2020cbj}, whereas the Monte Carlo generator provides the kinematic distributions of the nucleons entering the coalescence calculation~\cite{mahlein2023realistic}. 
Unlike the traditional \momCoal approach, coalescence is not determined by a sharp momentum threshold, but by the overlap between the proton--neutron pair and the antideuteron wave functions in phase space. 

An accurate modelling of the nucleus wave function is essential and several forms of the (anti)deuteron wave function are available in the literature~\cite{wiringa1995accurate, entem2017high, mahlein2023realistic}. 
The wave function used in this work is based on the \ArgonneVEighteen potential \cite{wiringa1995accurate}, constrained by \rm{pp} and \rm{np} inelastic scattering data, low-energy \rm{nn} scattering parameters, and deuteron binding energy.
Compared to the Gaussian approximation, preferred by many works because of the possibility to obtain analytical solutions for the quantum-mechanical problem despite not being realistic, the \ArgonneVEighteen wave function provides a data-constrained description of the deuteron internal structure. \ArgonneVEighteen was scrutinised in previous works and found to be effective in reproducing (anti)deuteron production in pp collisions at the LHC~\cite{mahlein2023realistic} as well as the \antideuteron yields measured by ALEPH in $\mathrm{e}^{+}\mathrm{e}^{-}$ collisions~\cite{DiMauro:2024kml}.
For these reasons, the implementation of the coalescence process within the quantum-mechanical approach follows the treatment in~\cite{mahlein2023realistic} and the \ArgonneVEighteen wavefunction is adopted as the default description of the antideuteron throughout this work. The only modification with respect to ~\cite{mahlein2023realistic} is the treatment of weak-decay chains through the equal-time prescription described below.

Once \antibHadron have been produced and decayed by \PYTHIA, all final-state antinucleons are processed by a quantum-mechanical coalescence afterburner. For each event, the algorithm proceeds via four stages: identification of candidate antiproton--antineutron pairs, construction of an equal-time configuration to enforce causality, evaluation of the pair kinematics in the pair rest frame, and calculation of the quantum-mechanical coalescence probability. Each step is described below.

First, all antinucleons in the final state are identified. For the discussion in Sec.~\ref{sec:Results}, these are distinguished between those originating directly from the decay of the \antibHadron hadron, referred to as prompt, and those produced through intermediate decays (i.e. from charmed hadrons or lighter hadrons) referred to as non-prompt.

Second, all antinucleons are then ordered according to their production time in the laboratory frame. 
This temporal ordering enforces the principle of causality, preventing the unphysical pairing of particles emitted at widely separated times or originating from causally disconnected regions of space-time.
If one antinucleon is produced earlier than the other, it is propagated to the production time of the later particle, thereby defining a common ``equal time'' for the pair.
This equal time is not an arbitrary reference time, but corresponds to the later of the two antinucleon production times.
The spatial separation and kinematic properties of the pair are then evaluated at this common time.
This prescription is particularly relevant for beauty-hadron decay chains, where antinucleons may be produced at different stages of the cascade.
Without this causal treatment, an antinucleon produced promptly in the \antibHadron decay could be artificially paired with another antinucleon produced much later, for instance, from the decay of a long-lived intermediate particle.
Such pairings would correspond to physically implausible coalescence configurations and may lead to spurious contributions.
For each \antiproton--\antineutron pair satisfying this equal-time construction, the total four-momentum is used to determine the velocity of the pair rest frame.
The space-time and momentum four-vectors of both antinucleons are then Lorentz-boosted into this frame.

Subsequently, the relative spatial separation $\Delta \vec{R}$ and relative momentum $\Delta \vec{P}$ are computed, and the quantity \mbox{$k^{*} = \frac{1}{2} \lvert \Delta \vec{P} \rvert$} is evaluated.
At this stage, we apply a kinematic selection requiring \mbox{${k^{*} < 0.2~\gevc}$}, to efficiently reject pairs unlikely to form a bound state~\cite{mihaylov2018femtoscopic, alice2018pp}.
The surviving pairs are retained as \antideuteron coalescence candidates.

Finally, their coalescence probability is evaluated from the overlap with the \ArgonneVEighteen-based Wigner distribution.
Unlike deterministic coalescence models, the final coalescence decision is therefore probabilistic.
The Wigner function is represented by a two-dimensional probability distribution in the $(\Delta R,\Delta P)$ phase space, where $\Delta R = |\Delta \vec{R}|$ and $\Delta P = |\Delta \vec{P}|$.
For each candidate pair, the calculated values of $\Delta R$ and $\Delta P$ are then tested against the \Argonne-based distribution to determine whether the coalescence condition is fulfilled.

Additional details on the Wigner function-based coalescence model employed in this work are reported in Appendix~\ref{sec:appendixb}.

\section{Results}
\label{sec:Results}
\begin{figure}[H]
  \centering
  \includegraphics[width=0.8\linewidth]{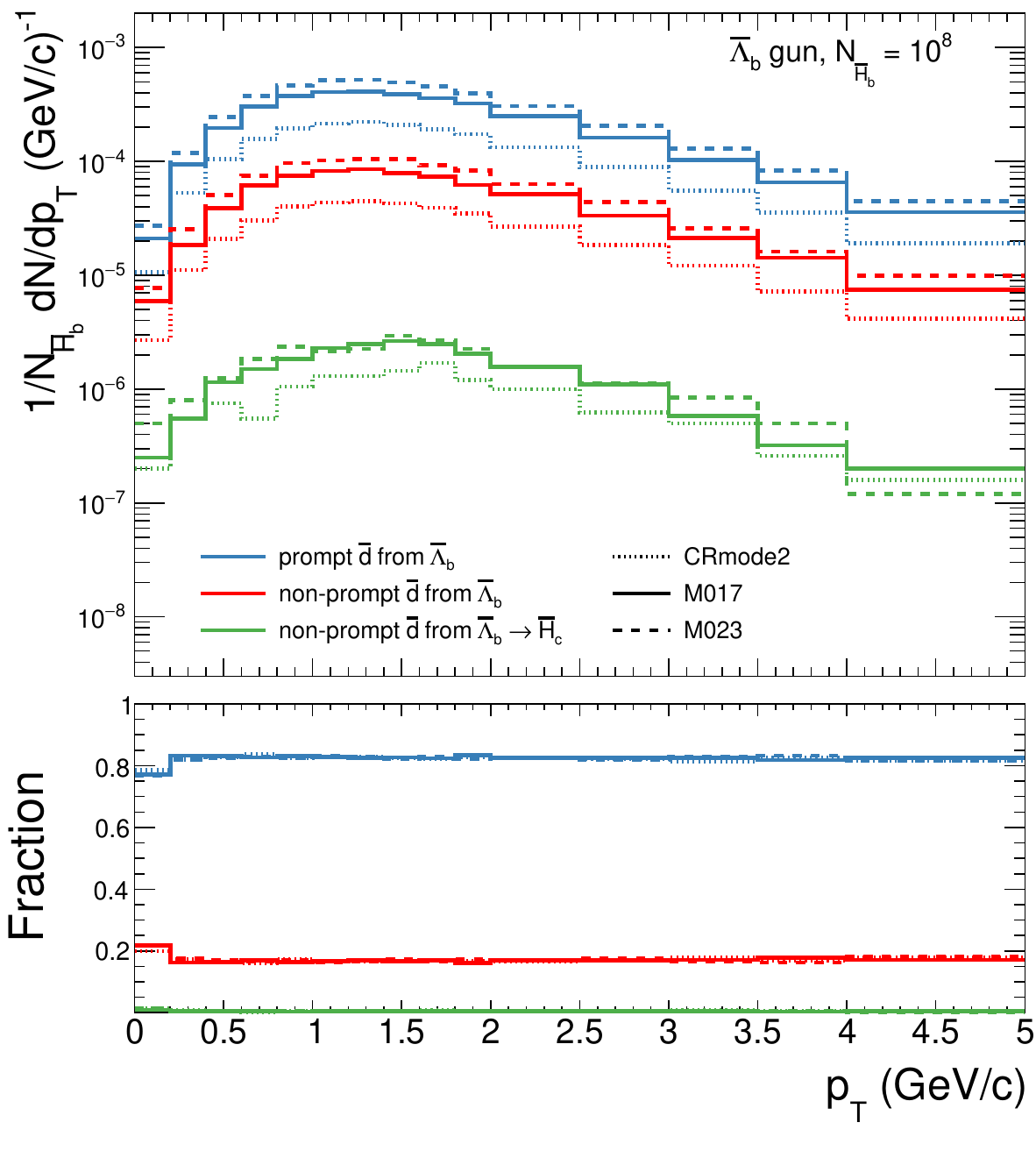}

  \caption{
    \textbf{Top:} Transverse-momentum spectrum of the antideuterons produced in the decay of \Lbbar.
    \textbf{Bottom:} Fraction of antideuterons produced from the \Lbbar decay.
    In both panels, the light-blue entries correspond to the prompt case.
    The red entries correspond to the non-prompt case, and the green entries isolate the non-prompt subset where the antinucleons originate from charm hadrons produced in the \Lbbar decay.}
  \label{fig:figure1}
\end{figure}

A comprehensive analysis of the decay chains of both \Lbbar and \Bminus is performed to answer the question of how the antideuterons are formed.
The total number of \antibHadron simulated is $10^8$.
The antideuterons produced at the end of the simulation chain are classified by tracing their ancestors back through the event record.
Two main production channels are distinguished: prompt antideuterons, shown in light-blue in \Fig{fig:figure1}, and non-prompt production.
The latter is further subdivided into two sub-categories.
In the first sub-category, at least one of the two antinucleons has a charm-hadron ancestor while neither of them has a strange-hadron ancestor.
This class of antideuterons is shown in green in \Fig{fig:figure1}.
The second sub-category collects all remaining non-prompt antideuterons, whose ancestors consist exclusively of light-quark hadrons; these are shown in red in \Fig{fig:figure1}.
The presence of strange-hadron ancestors was explicitly investigated: the decay chains were scanned for antinucleons whose earliest identifiable ancestor was a strange hadron with no charm content.
No \antideuteron satisfying this criterion was found in any of the three configurations studied, so we conclude that the strange channel does not contribute to the \antideuteron formation.
This would be naively expected, as the longer lifetime of strange hadrons producing nucleons would lead to a larger spatial separation from the original \antibHadron, thus reducing the coalescence probability.

To understand how the prediction depends on the hadronization configuration, the analysis is repeated for each of the three \PYTHIA tunes described in Sec.\ref{sec:pythia}: \firsttune, \secondtune and \thirdtune.
The tables listing the most frequent decay chains observed with the three configurations are provided in Appendix~\ref{sec:Appendix} for both \Lbbar and \Bminus.

Two main features emerge from the study of the decay chains. 
First, antideuteron production is dominated by prompt antinucleons. 
Second, the contribution from the charm-mediated decay does not exceed a few percent.

Figure~\ref{fig:figure1} summarizes the results for the \Lbbar case in terms of \antideuteron fractions and \pT-differential distribution.
Across the three tunes, the charm chains contribute only with $\approx 1\%$ to the total \antideuteron yield.
Furthermore, less than $20\%$ of the antideuterons are produced through other non-prompt channels, whereas the remaining majority originates promptly from the \Lbbar decay.
The total \antideuteron production increases passing from \firsttune to \secondtune, with the prompt component (which constitutes the dominant contribution) showing a slight enhancement.
In contrast, \thirdtune leads to a significant suppression, with the prompt yield reduced to approximately half of that observed in the other two cases.
For all three tunes, the \pT-distributions peak at around $1.0$--$1.5$~\gevc.
Although the absolute yields depend significantly on the hadronization model, the predicted \pT-distributions exhibit very similar shapes across all considered tunes.

\begin{figure}[H]
  \centering
  \includegraphics[width=0.8\linewidth]{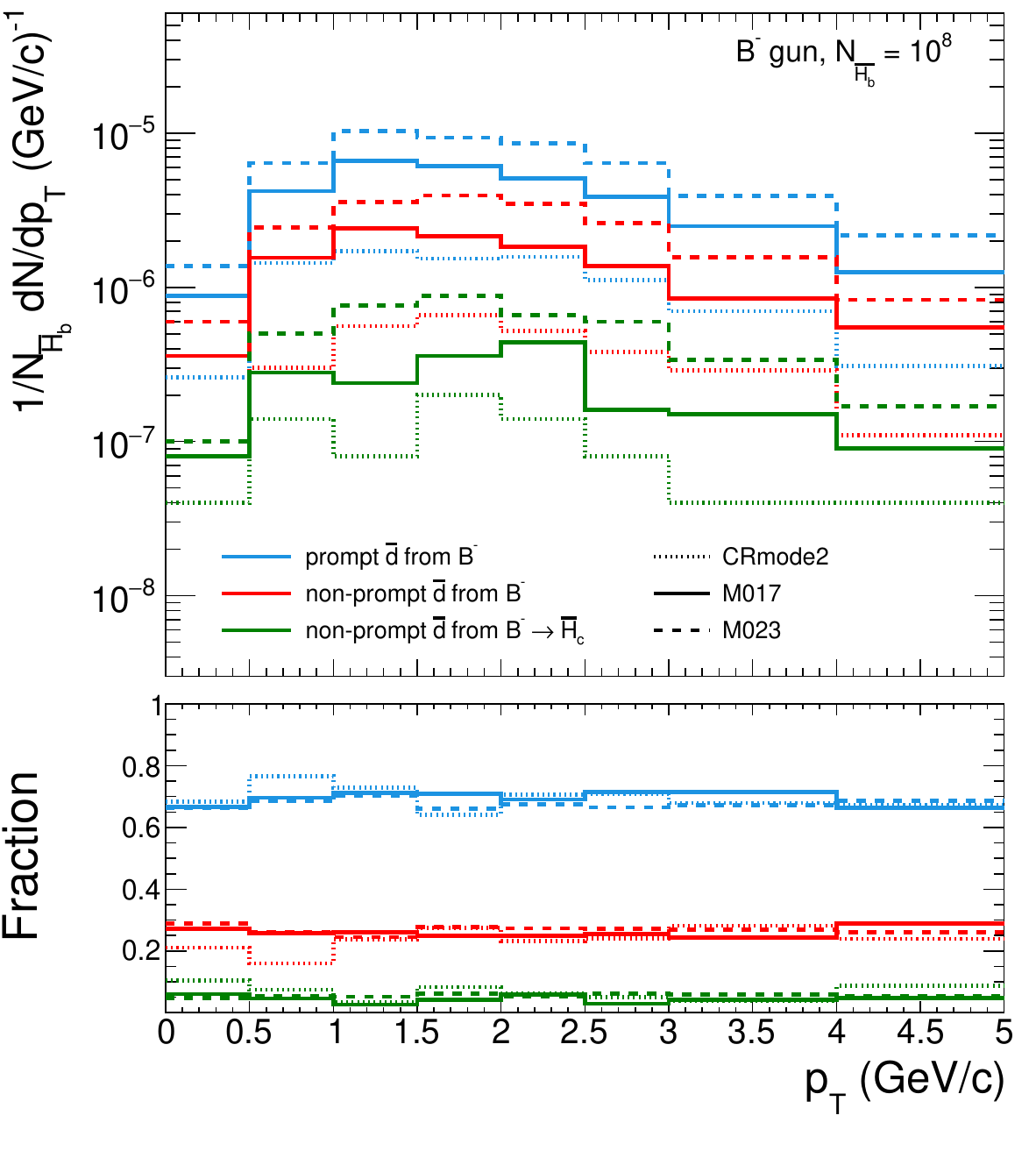}
  \caption{
    \textbf{Top:} Transverse-momentum spectrum of the antideuterons produced in the decay of \Bminus.
    \textbf{Bottom:} Fraction of antideuterons produced from the \Bminus decay.
    In both panels, the light-blue lines correspond to the prompt case.
    The red lines correspond to the non-prompt case, and the green lines isolate the non-prompt subset where the antinucleons originate from charm hadrons produced in the \Bminus decay.}
  \label{fig:bmeson}
\end{figure}

The same analysis was repeated for the \Bminus mesons. Although the absolute yield is smaller than for the \Lbbar, the overall production mechanism exhibits similar characteristics.
In \Fig{fig:bmeson}, the results for the \Bminus~case are reported with the same colour convention as in \Fig{fig:figure1}.
Approximately $3$--$4\%$ of
antideuterons originate from charm chains, while about $30\%$ belong to the non-prompt category.
For both cases, the fractions are \pT-independent in the considered range.
The majority of \antideuteron originating from the \Bminus is produced through prompt decay channels.
In general, the total antideuteron yield is lower than in the \Lbbar case, which explains the larger statistical uncertainties associated with the prediction.
Comparing the \pT distributions shows that the transition from the first two configurations to \thirdtune results in a suppression of the peak by approximately one order of magnitude, although the overall shape of the distribution is preserved.
In all cases, the distribution's peaks are in the \pT range $1.5$--$2.0~\gevc$.

The branching ratios for the decays under study were computed for all tunes and are reported in Table~\ref{tab:BR_antid}.
While varying \probQQtoQ does not change their order of magnitude, the branching ratios obtained with \thirdtune are, in both cases, about one order of magnitude smaller.
As the number of antideuterons is reduced in this last configuration, the statistical uncertainties increase accordingly.
The most favourable condition is obtained with the tune \secondtune in \PYTHIA that, as expected, corresponds to an increased number of antinucleons produced through the decays.
The spread among the three tunes should not be interpreted as a complete systematic uncertainty but rather as an estimate of the sensitivity of the prediction to the hadronization description.
\begin{table}[h]
  \centering
  \renewcommand{\arraystretch}{1.4}
  \begin{tabular}{lcc}
    \hline
                & \multicolumn{2}{c}{Branching Ratios}                                  \\
                & $\Lbbar \to \bar{\rm{d}} + X$        & $\Bminus \to \bar{\rm{d}} + X$ \\
    \hline
    \firsttune  & $(1.123\pm0.003)\times 10^{-3}$      & $(2.60\pm0.05)\times 10^{-5}$  \\
    \secondtune & $(1.408\pm0.004)\times 10^{-3}$      & $(4.34\pm0.07)\times 10^{-5}$  \\
    \thirdtune  & $(5.68\pm0.02)\times 10^{-4}$        & $(7.4\pm0.3)\times 10^{-6}$    \\
    \hline
  \end{tabular}
  \caption{Branching ratios for \antideuteron production from \Lbbar and \Bminus decays for the three configurations studied.}
  \label{tab:BR_antid}
\end{table}

The results presented above indicate that the predicted branching ratios are sufficiently large to motivate dedicated experimental searches.
We now turn to the experimentally relevant observables to investigate whether the predicted antideuterons populate the kinematic region accessible to the ALICE experiment at the LHC.

\begin{figure}[H]
  \centering
  \includegraphics[width=0.8\linewidth]{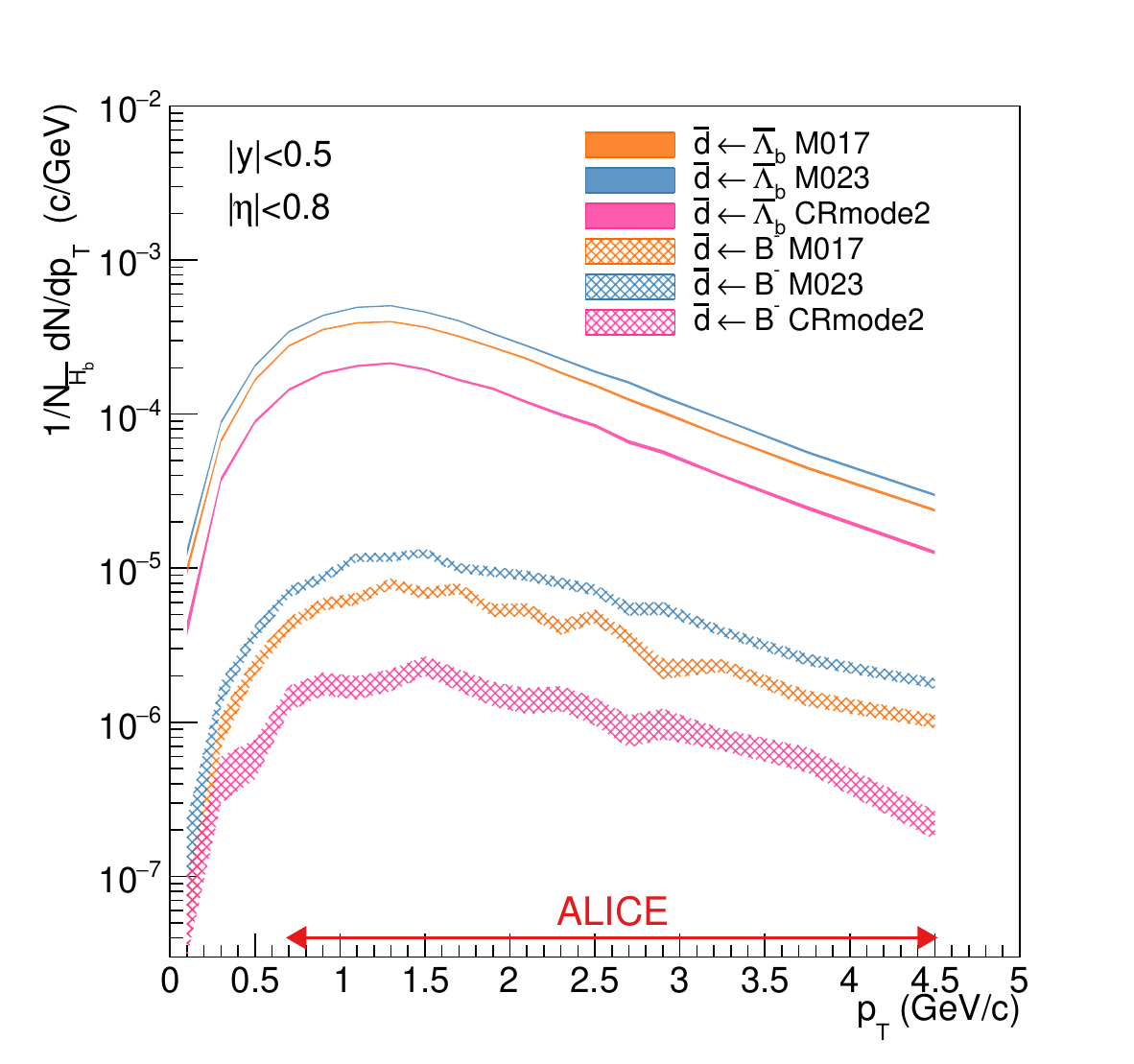}
  \caption{
    The light blue, orange and pink full lines represent the total transverse-momentum distribution of antideuterons produced from \Lbbar decays and the others (down) from the \Bminus for \firsttune, \secondtune and \thirdtune tunes.
    The red double-headed arrow highlights the transverse momentum range within which antideuterons are identifiable with ALICE.
    All \pT-distributions were obtained with a pseudorapidity selection for the antideuterons of $|\eta|<0.8$ and $|y|<0.5$ in rapidity. Bands represent statistical uncertainties.}
  \label{fig:totald}
\end{figure}

The transverse momentum distributions of all antideuterons produced from \Lbbar and \Bminus decays were extracted in the rapidity interval $|y|<0.5$ and pseudorapidity $|\eta|<0.8$.
This rapidity window was chosen to match the acceptance of the ALICE detector, which has already successfully measured prompt \antideuteron production in various collision systems~\cite{ALICE_dbar_PbPb2016, ALICE_dbar_pp2019, ALICE_dbar_rap2025, ALICE_dbar_resonance2025, ALICE:2020foi,ALICE:2022veq,ALICE:2022pbb}.
The results, reported in \Fig{fig:totald}, show that a larger overall contribution from \Lbbar can be observed compared to \Bminus, regardless of the tune.
The distributions exhibit a similar trend, with a maximum in the range $1 - 1.5~\gevc$ for the antideuterons produced from the \Lbbar and $1.5 - 2~\gevc$ for the antideuterons produced from \Bminus.
Within each particle, the choice of the tune affects the distribution by shifting it up or down as a normalisation factor; it does not modulate the \pT-dependence.

It has to be noted that these distributions and production yields must still be folded with the \antibHadron production cross-section and the relative transition rates in pp collisions, which, being larger for \Bminus than for \Lbbar~\cite{LHCb:2023wbo,navas2024review}, partially compensate for this hierarchy in the decay ratios.

The transverse momentum interval over which the ALICE detector \cite{ALICE:2022wpn} can identify deuterons and antideuterons spans approximately from $0.7$ to $4.5~\gevc$ in \pT and is indicated in the figure by a red double arrow.
The predicted antideuteron spectra peak within this interval for all generator configurations, indicating that the kinematic acceptance of ALICE overlaps well with the region where the signal is expected to be largest.
\section*{Conclusions}
\label{sec13}
In this study, \PYTHIA was used to simulate \Lbbar and \Bminus decay chains, starting from realistic input distributions and considering three different tunes to account for different hadronization scenarios.
A state-of-the-art quantum-mechanical coalescence model was employed to implement cluster formation, using the \ArgonneVEighteen potential to model the antideuteron wave function and implementing causality in the treatment of weak decays. 
Prompt and non-prompt antideuteron production were treated separately, with the contribution from the former found to be dominant. 
Among the non-prompt channels, the charm contribution was found to be negligible compared to the light-hadron contribution.

The branching ratios obtained for the three tune configurations lie in the intervals $(5.68 \pm 0.02)\times10^{-4}<\brLbtod<(1.408 \pm 0.004)\times10^{-3}$, 
and
$(7.4 \pm 0.3)\times10^{-6}<\brBtod<(4.34 \pm 0.07)\times10^{-5}$.

These ranges correspond to variations of approximately a factor of 2.5 and 6, respectively, across the considered PYTHIA tunes, providing an estimate of the current modeling uncertainty associated with the hadronization description.
Within the realistic simulation setup discussed in this work, \secondtune and \thirdtune define the upper- and lower-yield scenarios, respectively.

The predicted kinematic distributions indicate that the \antideuteron produced in beauty hadron decay populate the region where they can be identified with the ALICE detector, motivating dedicated experimental searches for these decay channels.

Overall, these results provide the first quantitative benchmark for experimental searches for antideuterons from beauty-hadron decays, together with a phenomenological framework that incorporates recent advances in beauty-hadron phenomenology and coalescence modeling. 
The study of antideuteron from beauty-hadron decays offers a new testing ground for investigating the universality of light (anti)nuclei formation mechanism across different production environments, with potential implications for indirect Dark Matter searches and, more generally, for cosmic-ray physics.


\section*{Declarations}
This work has been carried out thanks to the funding from the European Research Council (ERC) under the European Union’s Horizon 2020 research and innovation programme, grant agreement No. 950692 - CosmicAntiNuclei.
\clearpage
\appendix
\section{Appendix}
\label{sec:Appendix}
A comparison between the branching ratios (BR) into (anti)nucleons is reported for the three configurations used in \PYTHIA.
The total number of $\overline{\rm{H_b}}$ particles simulated is $10^7$. Only channels containing at least one antinucleon are shown.
\begin{table}[htbp]
  \centering
  \footnotesize
  \renewcommand{\arraystretch}{1.05}
  \begin{tabular}{l r r r}
    \toprule
    \multicolumn{1}{c}{Channel}                 &
    \multicolumn{1}{c}{\texttt{probQQtoQ=0.17}} &
    \multicolumn{1}{c}{\texttt{probQQtoQ=0.23}} &
    \multicolumn{1}{c}{\texttt{CRmode2}}                                      \\
    \midrule
    $\bar{p}$                                   & 6.58416 & 6.45555 & 6.60240 \\
    $\bar{n}$                                   & 6.12911 & 6.18870 & 6.14717 \\
    $\bar{n}\,p$                                & 2.78219 & 3.58143 & 2.84464 \\
    $\bar{n}\,n$                                & 1.26497 & 1.56237 & 1.29955 \\
    $\bar{p}\,p$                                & 1.16059 & 1.47552 & 1.17841 \\
    $\bar{p}\,n$                                & 0.40029 & 0.49721 & 0.41333 \\
    $\bar{p}\,\bar{n}\,p$                       & 0.23328 & 0.29794 & 0.24475 \\
    $\bar{n}\,\bar{n}\,p$                       & 0.09984 & 0.13061 & 0.10605 \\
    $\bar{p}\,\bar{n}\,n$                       & 0.10289 & 0.12631 & 0.10603 \\
    $\bar{p}\,\bar{p}\,p$                       & 0.09586 & 0.11709 & 0.09677 \\
    $\bar{p}\,\bar{n}$                          & 0.06094 & 0.07888 & 0.06141 \\
    $\bar{n}\,\bar{n}\,n$                       & 0.04803 & 0.06216 & 0.04983 \\
    $\bar{p}\,\bar{p}$                          & 0.04219 & 0.05346 & 0.04335 \\
    $\bar{p}\,\bar{p}\,n$                       & 0.04264 & 0.05157 & 0.04394 \\
    $\bar{n}\,\bar{n}$                          & 0.01092 & 0.01385 & 0.01137 \\
    $\bar{p}\,\bar{p}\,\bar{n}\,p\,p$           & 0.00173 & 0.00291 & 0.00180 \\
    $\bar{p}\,\bar{n}\,\bar{n}\,n\,p$           & 0.00165 & 0.00290 & 0.00174 \\
    $\bar{p}\,\bar{p}\,\bar{n}\,n\,p$           & 0.00156 & 0.00288 & 0.00167 \\
    $\bar{p}\,\bar{n}\,n\,p$                    & 0.00058 & 0.00100 & 0.00053 \\
    $\bar{p}\,\bar{p}\,\bar{n}\,p$              & 0.00045 & 0.00090 & 0.00036 \\
    $\bar{p}\,\bar{p}\,\bar{p}\,p\,p$           & 0.00052 & 0.00086 & 0.00052 \\
    $\bar{p}\,\bar{n}\,\bar{n}\,p\,p$           & 0.00053 & 0.00085 & 0.00054 \\
    $\bar{p}\,\bar{n}\,\bar{n}\,n\,n$           & 0.00050 & 0.00073 & 0.00046 \\
    $\bar{p}\,\bar{n}\,\bar{n}\,p$              & 0.00030 & 0.00052 & 0.00029 \\
    $\bar{p}\,\bar{n}\,p\,p$                    & 0.00039 & 0.00051 & 0.00028 \\
    $\bar{p}\,\bar{n}\,\bar{n}\,n$              & 0.00033 & 0.00047 & 0.00023 \\
    $\bar{p}\,\bar{p}\,\bar{n}\,n\,n$           & 0.00028 & 0.00044 & 0.00018 \\
    $\bar{p}\,\bar{p}\,\bar{p}\,n\,p$           & 0.00022 & 0.00043 & 0.00031 \\
    $\bar{p}\,\bar{p}\,p\,p$                    & 0.00026 & 0.00042 & 0.00022 \\
    $\bar{p}\,\bar{p}\,n\,p$                    & 0.00013 & 0.00040 & 0.00016 \\
    $\bar{p}\,\bar{p}\,\bar{n}\,n$              & 0.00018 & 0.00037 & 0.00028 \\
    $\bar{p}\,\bar{p}\,\bar{p}\,p$              & 0.00019 & 0.00027 & 0.00026 \\
    $\bar{p}\,\bar{n}\,n\,n$                    & 0.00018 & 0.00024 & 0.00017 \\
    $\bar{p}\,\bar{p}\,\bar{p}\,n$              & 0.00006 & 0.00007 & 0.00007 \\
    $\bar{p}\,\bar{p}\,\bar{p}\,n\,n$           & 0.00002 & 0.00005 & 0.00001 \\
    $\bar{p}\,\bar{p}\,n\,n$                    & 0.00005 & 0.00005 & 0.00009 \\
    $\bar{p}\,\bar{p}\,\bar{p}$                 & 0.00001 & 0.00004 & ---     \\
    $\bar{p}\,\bar{p}\,\bar{n}$                 & ---     & 0.00003 & 0.00003 \\
    $\bar{p}\,n\,p$                             & 0.00001 & 0.00003 & 0.00001 \\
    $\bar{p}\,n\,n$                             & 0.00001 & 0.00001 & 0.00002 \\
    $\bar{p}\,\bar{n}\,\bar{n}$                 & ---     & 0.00001 & ---     \\
    $\bar{p}\,p\,p$                             & 0.00001 & ---     & ---     \\
    \bottomrule
  \end{tabular}
  \caption{Branching ratios (\%) of \Lbbar decays to channels containing at least one
antinucleon, for three different generator configurations.
The listed channels are inclusive in additional particles, implicitly denoted by $+\rm{X}$.}
  \label{tab:br_antinucleon}
\end{table}

\begin{table}[htbp]
  \centering
  \renewcommand{\arraystretch}{1.04}
  \begin{tabular}{l r r r}
    \toprule
    \multicolumn{1}{c}{Channel}                 &
    \multicolumn{1}{c}{\texttt{probQQtoQ=0.17}} &
    \multicolumn{1}{c}{\texttt{probQQtoQ=0.23}} &
    \multicolumn{1}{c}{\texttt{CRmode2}}                                      \\
    \midrule
    $\bar{p}\,n$                                & 1.78264 & 2.25676 & 0.95737 \\
    $\bar{p}$                                   & 1.16802 & 1.34801 & 0.86987 \\
    $\bar{n}\,n$                                & 0.77588 & 0.95300 & 0.42539 \\
    $\bar{p}\,p$                                & 0.73544 & 0.92773 & 0.38254 \\
    $\bar{n}$                                   & 0.26443 & 0.33324 & 0.13518 \\
    $\bar{n}\,p$                                & 0.25469 & 0.31180 & 0.14261 \\
    $\bar{p}\,\bar{n}\,n\,p$                    & 0.00188 & 0.00319 & 0.00050 \\
    $\bar{p}\,\bar{p}\,n\,p$                    & 0.00131 & 0.00204 & 0.00028 \\
    $\bar{p}\,\bar{n}\,n\,n$                    & 0.00113 & 0.00202 & 0.00025 \\
    $\bar{p}\,\bar{n}\,n$                       & 0.00097 & 0.00171 & 0.00019 \\
    $\bar{p}\,\bar{p}\,p\,p$                    & 0.00095 & 0.00171 & 0.00019 \\
    $\bar{p}\,\bar{p}\,p$                       & 0.00081 & 0.00142 & 0.00020 \\
    $\bar{p}\,\bar{n}\,p$                       & 0.00089 & 0.00127 & 0.00018 \\
    $\bar{p}\,\bar{n}\,p\,p$                    & 0.00074 & 0.00102 & 0.00020 \\
    $\bar{p}\,\bar{p}\,n$                       & 0.00050 & 0.00087 & 0.00011 \\
    $\bar{n}\,\bar{n}\,n\,n$                    & 0.00043 & 0.00086 & 0.00008 \\
    $\bar{p}\,n\,p$                             & 0.00057 & 0.00077 & 0.00013 \\
    $\bar{p}\,p\,p$                             & 0.00046 & 0.00069 & 0.00008 \\
    $\bar{n}\,\bar{n}\,n\,p$                    & 0.00056 & 0.00067 & 0.00015 \\
    $\bar{n}\,n\,p$                             & 0.00039 & 0.00059 & 0.00005 \\
    $\bar{p}\,\bar{p}\,n\,n$                    & 0.00024 & 0.00057 & 0.00014 \\
    $\bar{n}\,\bar{n}\,n$                       & 0.00033 & 0.00053 & 0.00012 \\
    $\bar{p}\,n\,n$                             & 0.00028 & 0.00042 & 0.00008 \\
    $\bar{n}\,n\,n$                             & 0.00022 & 0.00032 & 0.00004 \\
    $\bar{p}\,\bar{n}$                          & 0.00017 & 0.00031 & 0.00003 \\
    $\bar{n}\,\bar{n}\,p$                       & 0.00013 & 0.00030 & 0.00005 \\
    $\bar{n}\,\bar{n}\,p\,p$                    & 0.00017 & 0.00021 & 0.00003 \\
    $\bar{p}\,\bar{p}$                          & 0.00010 & 0.00018 & 0.00002 \\
    $\bar{n}\,p\,p$                             & 0.00015 & 0.00015 & 0.00003 \\
    $\bar{n}\,\bar{n}$                          & 0.00006 & 0.00005 & 0.00001 \\
    \bottomrule
  \end{tabular}
  \caption{Branching ratios (\%) of \Bminus decays to channels containing at least one
antinucleon, for three different generator configurations.
The listed channels are inclusive in additional particles, implicitly denoted by $+\rm{X}$.}
  \label{tab:br_antinucleon_Bminus}
\end{table}
\clearpage
The decay chains leading to the production of antinucleons that form the obtained antideuterons are shown below in \Tab{tab:decay_chains}.
\begin{table}[htbp]
  \centering
  \small
  \renewcommand{\arraystretch}{1.04}
  \begin{tabular}{l l}
    \toprule
    \multicolumn{1}{c}{\Lbbar decay chains}               &
    \multicolumn{1}{c}{\Bminus decay chains}                                                                       \\
    \midrule
    $\Lbbar \to \bar{p}$                                  & $\Bminus \to \bar{p}$                                  \\
    $\Lbbar \to \bar{n}$                                  & $\Bminus \to \bar{n}$                                  \\
    $\Lbbar \to \eta_c        \to \bar{n}$                & $\Bminus \to \eta_c        \to \bar{p}$                \\
    $\Lbbar \to J/\psi        \to \bar{n}$                & $\Bminus \to \eta_c        \to \bar{n}$                \\
    $\Lbbar \to J/\psi        \to \bar{p}$                & $\Bminus \to \eta_c(2S)    \to \bar{p}$                \\
    $\Lbbar \to \bar{\Delta}^{+}  \to \bar{n}$            & $\Bminus \to \eta_c(2S)    \to \bar{n}$                \\
    $\Lbbar \to \bar{\Delta}^{0}  \to \bar{n}$            & $\Bminus \to J/\psi        \to \bar{n}$                \\
    $\Lbbar \to \bar{\Delta}^{-}  \to \bar{n}$            & $\Bminus \to \bar{\Delta}^{+}  \to \bar{n}$            \\
    $\Lbbar \to \bar{\Delta}^{-}  \to \bar{p}$            & $\Bminus \to \bar{\Delta}^{0}  \to \bar{n}$            \\
    $\Lbbar \to \bar{\Delta}^{--} \to \bar{p}$            & $\Bminus \to \bar{\Delta}^{0}  \to \bar{p}$            \\
    $\Lbbar \to \bar{\Delta}^{0}  \to \bar{p}$            & $\Bminus \to \bar{\Delta}^{-}  \to \bar{n}$            \\
    $\Lbbar \to \eta_c \to \bar{\Delta}^{--} \to \bar{p}$ & $\Bminus \to \bar{\Delta}^{-}  \to \bar{p}$            \\
    $\Lbbar \to \eta_c \to \bar{\Delta}^{-}  \to \bar{p}$ & $\Bminus \to \bar{\Delta}^{--} \to \bar{p}$            \\
    $\Lbbar \to \eta_c \to \bar{\Delta}^{0}  \to \bar{n}$ & $\Bminus \to \eta_c \to \bar{\Delta}^{+}  \to \bar{n}$ \\
    $\Lbbar \to \eta_c \to \bar{\Delta}^{+}  \to \bar{n}$ & $\Bminus \to \eta_c \to \bar{\Delta}^{0}  \to \bar{n}$ \\
    $\Lbbar \to \eta_c \to \bar{\Delta}^{0}  \to \bar{p}$ & $\Bminus \to \eta_c \to \bar{\Delta}^{0}  \to \bar{p}$ \\
    $\Lbbar \to \eta_c \to \bar{\Delta}^{-}  \to \bar{n}$ & $\Bminus \to \eta_c \to \bar{\Delta}^{-}  \to \bar{n}$ \\
    $\Lbbar \to J/\psi \to \bar{\Delta}^{0}  \to \bar{n}$ & $\Bminus \to \eta_c \to \bar{\Delta}^{-}  \to \bar{p}$ \\
    $\Lbbar \to J/\psi \to \bar{\Delta}^{+}  \to \bar{n}$ & $\Bminus \to \eta_c \to \bar{\Delta}^{--} \to \bar{p}$ \\
                                                          & $\Bminus \to J/\psi \to \bar{\Delta}^{+}  \to \bar{n}$ \\
    \bottomrule
  \end{tabular}
  \caption{Decay chains through which the antinucleons forming the \antideuteron are produced, for the \Lbbar (left) and \Bminus (right) cases.
    The three generator configurations produce the same chains, differing only in their relative frequencies.
    Only the particles relevant to the study are shown; the term $ \rm{\;+\; X}$ has been omitted in the intermediate decays.}
  \label{tab:decay_chains}
\end{table}

\section{Appendix}
\label{sec:appendixb}
\subsection{Wigner function formalism}
A quantum-mechanical treatment of the coalescence process can be performed through the Wigner formalism, which accounts for the momentum distribution of the nucleons, the nucleus wave function, and the nucleon emitting source~\cite{Bellini:2019zqc, Blum:2017qnn, Bellini:2020cbj, kachelriess2020alternative}. 
Using this formalism, the differential $\overline{\rm d}$ spectrum is obtained by projecting the two-antinucleon density matrix $\rho_{\mathrm{pn}}$ onto the $\overline{\rm d}$ density matrix $\rho_{\mathrm{d}}$
$$
\frac{\mathrm{d}^3 N_{\mathrm{d}}}{\mathrm{d} p_{\mathrm{d}}^3}=\operatorname{tr}\left(\rho_{\mathrm{d}} \rho_{\mathrm{pn}}\right).
$$
The density matrices are defined as $\rho_{\mathrm{d}}=\left|\phi_{\mathrm{d}}\right\rangle\left\langle\phi_{\mathrm{d}}\right|$ and $\rho_{\mathrm{pn}}=\left|\psi_{\mathrm{pn}}\right\rangle\left\langle\psi_{\mathrm{pn}}\right|$,
where $\phi_{\mathrm{d}}$ and $\psi_{\mathrm{pn}}$ are the antideuteron wave function and the two-particle $\rm{\bar{p}-\bar{n}}$ wave function, respectively. By factorizing the spatial ($\vec{r}$) and momentum ($\vec{p}$) dependences of the wave function $\phi_{\mathrm{d}}$, one can write
$\phi_{\mathrm{d}}\left({\vec{r}}, {\vec{p}}\right) \propto \varphi_{\mathrm{d}} ({\vec{r}}) e^{i {\vec{p}} \cdot {\vec{r}}_{\mathrm{d}}},
$
where $\varphi_{\mathrm{d}}$ is the internal $\overline{\rm d}$ wave function, $\vec{r}_{\mathrm{d}}$ is the space-time position of the antideuteron and $\vec{p}$ its four-momentum. Taking this into account, the antideuteron spectrum takes the form \cite{Scheibl:1998tk,kachelriess2020alternative}:
\begin{equation}
\label{eq:wignerDspectrum}
\frac{\mathrm{d}^3 N_{\mathrm{d}}}{\mathrm{d} p_{\mathrm{d}}^3}= S \int \frac{\mathrm{d}^3 r_{\mathrm{d}} \mathrm{d}^3 r \mathrm{~d}^3 q}{(2 \pi)^6} \cdot  \mathcal{D}(\vec{r}, \vec{q})
\cdot W_{\mathrm{pn}}\left(\vec{p}_{\mathrm{d}} / 2+\vec{q}, \vec{p}_{\mathrm{d}} / 2-\vec{q}, \vec{r}_{\mathrm{p}}, \vec{r}_{\mathrm{n}}\right),
\end{equation}
where $S$ is a factor accounting for spin and isospin statistics, equal to $3 / 8$ for a $\overline{\rm d}$. $\vec{r}_{\mathrm{p}}$ and $\vec{r}_{\mathrm{n}}$ are the antiproton and antineutron positions, $\vec{r} \equiv \vec{r}_{\mathrm{p}}-\vec{r}_{\mathrm{n}}$, and $\vec{q}$ is defined as $\vec{q}\equiv\left(\vec{p}_{\mathrm{p}}-\vec{p}_{\mathrm{n}}\right) / 2 = \Delta \vec{p}/2$, where $\vec{p}_{\mathrm{p}}$ and $\vec{p}_{\mathrm{n}}$ are the antiproton and antineutron momenta, respectively.

The Wigner function of the $\rm{\bar{p}-\bar{n}}$ state is denoted by $W_{\mathrm{pn}}$, while $\mathcal{D}$ is the Wigner function of the antideuteron, defined as \cite{Scheibl:1998tk}:
$$
\mathcal{D}(\vec{r}, \vec{q})=\int \mathrm{d}^3 \xi e^{-i \vec{q} \cdot \vec{\xi}} \varphi_{\mathrm{d}}(\vec{r}+\vec{\xi} / 2) \varphi_{\mathrm{d}}^*(\vec{r}-\vec{\xi} / 2) .
$$
The antideuteron Wigner function is normalized such that:
\begin{equation}
\label{eq:Wignormalized}
\int d^3r \int \frac{d^3q}{(2 \pi)^3}\mathcal{D}(\vec{r}, \vec{q}) = 1 .
\end{equation}

The choice of the $\overline{\rm d}$ wave function $\varphi_{\mathrm{d}}$ affects the form of the antideuteron Wigner function $\mathcal{D}(\vec{r}, \vec{q})$.
Concerning the $\rm{\bar{p}-\bar{n}}$ Wigner function $W_{\mathrm{pn}}$, the spatial and momentum dependences are factorized as:
$$
W_{\mathrm{pn}}=H_{\mathrm{pn}}\left(\vec{r}_{\mathrm{p}}, \vec{r}_{\mathrm{n}}\right) G_{\mathrm{pn}}\left(\vec{p}_{\mathrm{d}} / 2+\vec{q}, \vec{p}_{\mathrm{d}} / 2-\vec{q}\right),
$$
where $G_{\mathrm{pn}}$ is the two-particle momentum distribution, obtained from a \texttt{PYTHIA} modeling of the $\bar{p}$ and $\bar{n}$ production processes, including both the single-particle momentum distributions of the nucleons and their correlations.
For the spatial term $H_{\mathrm{pn}}$, the $\bar{n}$ and $\bar{p}$ distributions can be separated:
$$
H_{\mathrm{pn}}\left(\vec{r}_{\mathrm{p}}, \vec{r}_{\mathrm{n}}\right)=h\left(\vec{r}_{\mathrm{p}}\right) h\left(\vec{r}_{\mathrm{n}}\right),
$$
where $h\left(\vec{r}_{\mathrm{p}}\right)$ and $h\left(\vec{r}_{\mathrm{n}}\right)$ are the spatial single-particle distributions, obtained from the same \texttt{PYTHIA} modeling.
It is noteworthy that in applications of this model to pp collisions, such as in the one reported in \cite{mahlein2023realistic}, the nucleon emitting source can be further tuned or constrained based on femtoscopic measurements~\cite{sourceSizeHMpp}. In this work, the same does not apply since only nucleons stemming from a beauty particle decay chain are used for coalescence.

As last ingredient, the Argonne $v_{18}$ deuteron wave function is used for the reasons discussed in Sec.~\ref{sec:coalescence}. This wave function is derived from the Argonne $v_{18}$ potential, a phenomenological potential for the deuteron tuned to $\rm{pp}$ and $\rm{np}$ inelastic scattering data, low-energy $\rm{nn}$ scattering parameters, and the deuteron binding energy \cite{wiringa1995accurate}. It is characterised by a total of $18$ parameters: $14$ charge-independent terms, $3$ charge-independence breaking terms, and $1$ charge-asymmetry term.
In this potential, the deuteron wave function has the form:

\begin{equation}
    \varphi_{\mathrm{d}}(\vec{r})=\frac{1}{\sqrt{4 \pi} r}u(r) \chi_{1 m}
\end{equation}
where $\chi_{1 m}$ is a spinor, and $u(r)$ is the radial $\mathrm{S}$-wave component.
The wave function is normalized as follows:
$$
\int \mathrm{d}^3 r\left|\varphi_{\mathrm{d}}(\vec{r})\right|^2=\int \mathrm{d}^3 r \frac{1}{4 \pi r^2}u^2(r)=1 .
$$
In the applied coalescence afterburner, we used the Argonne-derived Wigner function $\mathcal{D}(r,q)$ tabulated as a function of $r$ and $q$ as employed in  \cite{mahlein2023realistic}.

\backmatter
\bibliography{bibliography}

\end{document}